\documentclass{article}[12pt]
\usepackage[T1,T2A]{fontenc}
\usepackage{amsmath,amssymb}
\usepackage{graphicx}%
\usepackage{verbatim}
\usepackage{xcolor}

\newtheorem{assume}{Assumption}
\def\stackunder#1#2{\mathrel{\mathop{#2}\limits_{#1}}}

\newcommand{\Lee}[1]{\stackunder{#1}{\rm L}}

\newcounter{Ris}
\newcommand{\order}[3]{#1^{(#2)}\ \!#3}
\def\stackunder#1#2{\mathrel{\mathop{#2}\limits_{#1}}}
\newcommand{\pert}[2]{\stackrel{(#2)}{#1}}

\begin{document}

\begin{center}
{\bf \Large Yu. G. Ignat'ev\footnote{The Institute Of Physics, The N.I. Lobachevsky Institute of Mathematics and Mechanics of the Kazan Federal University, 420008 Russia, Kazan, str. Kremlevskaya 18; email Ignatev-Yurii@mail.ru}} \\[12pt]
{\bf \Large The Method of Self-Consistent Field and Macroscopical Einstein Equations for the Early Universe} \\[12pt]
\end{center}

\abstract{The complete theory of cosmological evolution's macroscopic description is constructed with a help of self-consistent  field. The theory includes a subsystem of linear equations of perturbations' evolution and nonlinear macroscopic Einstein equations and equations of a scalar field. We provide examples of system's solution which show the principal difference of the cosmological models of the early Universe built on homogenous and locally fluctuating scalar fields. 
}

\section*{Introduction}
The main statements of the macroscopic theory of gravitation based on a method of self-consistent field were formulated in papers \cite{stfi15,G&C16,stfi16}. Along with that, the macroscopic Einstein equations for the cosmological model with $\Lambda$ - term in a second order of perturbations' theory over transverse gravitation perturbations were obtained and integrated. The Universe, which was practically empty, filled in with gravitational radiation was considered in \cite{stfi15}, \cite{G&C16}. Scalar gravitational perturbations cannot exist in such Universe since the existence of matter is required for it. More precisely, scalar gravitational perturbations can only appear in a second order of perturbations' theory in a capacity of perturbations of average density of gravitational radiation's energy . The paper \cite{stfi16} considered a cosmological model with a scalar field however only transverse perturbations of metric which corresponded to gravitational waves were accounted. Such perturbations do not lead to perturbations of scalar field in a linear approximation, therefore such a model is to a great extent identical to the model of the empty Universe filled by gravitational radiation. In this paper we will consider a complete macroscopical model of the Universe with the account of both longitudinal (scalar) and vector perturbations of metric. We will consider classical scalar field in a capacity of matter\footnote{This work was funded by the subsidy allocated to Kazan Federal University for the state assignment in the sphere of scientific activities.}.

\section{The Self-Consistent Statistical Approach to Description of Metric's Local Fluctuations}

\subsection{The Self-Consistent Field Method}

Statistical theory of obtaining {\it macroscopic Einstein equations} can be developed in a similar way to many particle theory within the
framework of the approach of {\it self-consistent field}, which initially originated in celestial mechanics and then was applied in many
particle theory (see well-known classical papers P. Weiss, 1907; I. Langmuir, 1913;
L. Thomas, 1927; E. Fermi, 1928; D. Hartree, 1928;
V.\,A. Fok, 1930). A.A.Vlasov played a specific role in the development of the self-consistent field's method by carrying out an in-depth
analysis of physical properties of charged particles of plasma, showing inapplicability of the Boltzmann's gas-kinetic equation to
description of plasma and suggestion of a new kinetic plasma equation (Vlasov equation) describing collective interaction of plasma
particles by means of the self-consistent field (see \cite{Vlasov1938} (1938), \cite{Vlasov1945}, \cite{Vlasov}).
Vlasov's theory was refined by L.\,D.Landau in  \cite{Land_Damp} (1946) , complete proved and generalized by N.\,N.Bogolyubov
\cite{Bogolyubov46} (1946) and brilliantly applied to quantum statistics and superfluidity theory \cite{Bogolyubov47} (1947).
Let us note the classical monograph by Chandrasekhar \cite{Chandr} (1942) where, on the basis of the
self-consistent eld method, the principles of stellar dynamics were formulated, and, in essence, the theory of galactic
structures was built.

According to the self-consistent field method, a separate particle's motion can be described as a motion in an aggregate averaged field of
particles of the rest of the system, neglecting the influence of this single particle on the system's dynamics. The conditions of
applicability of the method are long-range character of particle interactions and  large number of interacting particles $N\gg1$. Similarly,
we can consider the method with regard to pure nonlinear field systems. Here, a separate small field mode, that is characterized by certain
field degrees of freedom, acts as a single particle, and the condition of the method applicability is a long-range character of the field
and a large number of its degrees of freedom. The gravitational interaction perfectly corresponds to these conditions: the conservation law
of the full energy-mass and the absence of negative <<gravitational charges>> ensures its long-range character,while a large number of
degrees of freedom is laid in the field nature of interaction itself. Therefore we can consider systems with the gravitational interaction
by the self-consistent field method, where each microscopic mode of gravitational perturbations is small whereas the macroscopic self-consistent gravitational field is large.

Let us notice that the model based on quantum-field statistics would be more complete macroscopic model of the early Universe.
Quantum-gravitational effects in the early Universe are essential for wavelengths comparable with the Universe's curvature
radius $r_k=\sigma^{-1/4}$:
\[
\sigma= \sqrt{R_{ijkl}R^{ijkl}}=6H^4(1+w^2),
\]
where $H$ is a Hubble constant, $w$ is invariant cosmological acceleration. Thus, the quantum effects are important for field modes with the wave vector
\[
k^4\sim 6H^4(1+w^2).
\]
This condition for the inflationary stage of the Universe $H=\Lambda$, $w=1$, where $\Lambda$ is a cosmological constant, takes the following form:
\[
k\sim \Lambda.
\]
Since $k\sim n/a(t)=n \exp(-\Lambda t)$, the last condition takes the form $n\sim \Lambda\exp(\Lambda t)$. As a result of this condition, massless particles with fixed mode $n(t)$ are generated in each time instant, which in the end leads to the known flat spectrum of generated gravitons  \cite{starob1} (see also \cite{starob2}).

The reverse reaction on background geometry of the early Universe in form of massless particles's generation  was discussed in \cite{back1} -- \cite{back4} (2014 -- 2016) and other works. Let us notice that these works were based on semiphenomenological approach in which particles' generation is described by thermal spectrum with a temperature defined by the curvature. The cited papers consider stability of the classic background metric with respect to reverse reaction's factor and different authors arrive to directly opposed conclusions. In particular, in \cite{back3} it is noted that conclusions of stability of the background metric are based on a scheme not containing averaging operation and therefore, are incorrect. Let us also notice that reverse influence of perturbed metrics on particles' generation process was not accounted in the cited papers, hence the considered models are not self-consistent and account only unitransverse connection.

Thus it can be stated that the problem of quantum-statical description of the macroscopic early Universe in quantum cosmology\footnote{Let us notice that such description should be based on density matrix, however corresponding mathematical apparatus for quantum gravitation has not been developed.} has not been investigated sufficiently enough. In the given paper we will be developing the classical theory of the macroscopic Universe.  Let us notice that there is a semiclassical bridge connecting classical statistical description of the early Universe with the quantum theory. In the classical case random perturbations of metrics are defined by arbitrary initial conditions while in quantum case these perturbations are defined by instability of vacuum state and have determined character including sharply defined energy spectrum.
This energy spectrum should complement the classical theory, eliminating its arbitrariness, and, thereby, playing a role of aforementioned connection between quantum and classical theory.

\subsection{Averaging of Local Metric Fluctuations}

Let us suppose that the exact microscopic metric of the Riemannian space-time $V_4$ can be written in the form \cite{Bogolyub,Yubook1}:
\begin{equation}\label{gik_micro}
g_{ik}(x)=\overline{g_{ik}(x)}+\delta g_{ik}(x),
\end{equation}
where $\overline{g}_{ik}(x)$ is a certain average macroscopic metric corresponding to the macroscopic space-time
$\bar{V}_4$, and $\delta g_{ik}(x)$ are microscopic fluctuations of the metric, so that
\begin{equation}\label{dg<<1}
\overline{\delta g_{ik}\delta g^{jk}}\ll 1,
\end{equation}
and
\begin{equation}\label{bar_dg=0}
\overline{\delta g_{ik}}=0.
\end{equation}
Let us use the overline to designate (here and further) a certain operation of averaging of the metric
and the corresponding quantities, not yet defining it specifically.Let us just notice that this operation
is quite a delicate procedure, significantly depending on the method of measurement (see details in  \cite{Bogolyub,Yubook1}). Let us also notice that <<pioneer>> methods of statistical averaging of metrics being used by some researchers and, in essence, representing
a transition of methods of classical averaging by means of integrating the metric quantities over a spatial volume are obviously inapplicable in relativistic gravitation. First, as is known from Riemannian geometry, integration of a tensor over a volume is an ambiguous operation, and the result is not a tensor quantity. Second, since synchronization of observations is only
possible in a synchronized frame of reference, the physical meaning of the result of integration of the metric tenor over a volume cannot be determined
as well. Third, a macroscopic tool measuring the Universe's metric is simply non-realizable. Thus,
following the methods developed in \cite{Bogolyub,Yubook1}, we average the metrics over certain arbitrary quantities,
for instance, wave vectors, oscillations phases and other. For example, in the papers cited above, the metric was generated by massive particles,
and the averaging was carried out over coordinates of these particles, not being arguments of tensor fields. Generally speaking, correct metrics averaging as well as averaging of other tensor fields can be performed in the following way. The general theorem, well-known from the functional analysis states that a function in a metrics space can be expanded over {\it full} orthogonal set of eigenfunctions of linear self-conjugate operator of this space. Such an operator in this space is the D'Alembert operator. Therefore we can define this set by eigenfunctions of the D'Alembert operator being a self-conjugated linear operator on Riemann space $V_4$, corresponding to eigenvalues $\lambda$, in the following way:
\begin{equation}\label{self_func}
\overline{\triangle}\psi_\lambda =\lambda\psi_\lambda,
\end{equation}
where
\[\overline{\triangle}=\overline{\nabla}_i\overline{\nabla}\ ^i=\frac{1}{\sqrt{\overline{g}}}\frac{\partial}{\partial x^i}\sqrt{\overline{g}}\ \overline{g}\ ^{ik}\frac{\partial}{\partial x^k}\]
-- is a D'Alembert operator defined on metrics $\overline{g}_{ik}$. Let us notice that eigenfunctions
$\psi_\lambda$ are defined by certain {\it state vector} $\vec{\lambda}=[\ell_1,\ell_2,\ldots,\ell_n]$, over which the statistical averaging should be carried out. Let us notice that such averaging procedure is totally identical to averaging in the quantum field theory. Here we should impose certain requirements on the {\it distribution function} $f(x^i,\vec{\lambda})$, that should reflect the macroscopic properties of symmetry of the tensor fields being averaged.

We should understand that along with metric's fluctuations there exist corresponding small fluctuations of physical fields,
$\delta \phi_a$\footnote{In the considered case this is a single field $\phi_a=\Phi(x)$, however the above said is valid for any number of physical fields of any nature.},
\begin{equation}\label{phi_micro}
\phi_a=\overline{\phi}_a+\delta \phi_a,
\end{equation}
so that
\begin{equation}\label{dphi<<phi}
\overline{\delta \phi_a\delta \phi_a}\ll \overline{\phi^2_a},
\end{equation}
and
\begin{equation}\label{bar_df=0}
\overline{\delta\phi}_a=0.
\end{equation}
Averaging the Einstein equations with a help of similar procedure, let us find the following:
\begin{equation}\label{bar_einst_eq}
\overline{G^i_k(g)}=8\pi\overline{T^i_k(g,\phi_a)}+\Lambda \delta^i_k.
\end{equation}
As a consequence of nonlinearity of the Einstein tensor and tensor of the energy - momentum it is:
\begin{equation}\label{non_average}
\overline{G^i_k(g)}\not=G^i_k(\overline{g}); \quad \overline{T^i_k(g,\phi_a)}\not=T^i_k(\overline{g},\overline{\phi_a}),
\end{equation}
i.e., the a macroscopic source does not correspond to a macroscopic metric, as it turned to be the case in electrodynamics.

How should we obtain the equations that would describe the macroscopic metric in this case? The general principles of averaging of microscopic metrics
in a WKB approximation and at derivation of macroscopic Einstein equations were stated in the papers by Isaacson \cite{Isaakson1,Isaakson2} (1968), and later
they were used for construction of the theory of relativistic statistical systems with gravitational interaction in the author's papers \cite{Bogolyub0} (1983), \cite{Bogolyub} (2007), see also monographs \cite{Yubook1,Yubook2} (2010,2013). In \cite{Yu_Ark1,Yu_astr} (1990) this theory was applied to obtaining of the kinetic equation for photons spreading in a gravitationally fluctuating Friedmann's world where local fluctuations were originated by massive  <<particles>>. In particular, in \cite{Yu_Ark1,Yu_astr} (1989-1990) and conclusively in \cite{Yu_GC4_19} it was shown that accounting of such spherically symmetrical local fluctuations of gravitational field is equivalent to addition of the fluid with inflation state equation ($\varepsilon+p=0$) to Friedmann dust or addition of the  $\Lambda$-term  to Einstein equations. Let us also notice the work by A.V.Zakharov \cite{Zaharov} which is close to the subject, dedicated to obtainment of the kinetic equations for massive particles with account of pairwise gravitational interactions in the Friedman's world but not accounting the reverse influence of massive particles on the background metrics. Exactly this influence was taken into account in \cite{Yu_Ark0,Yu_Ark1,Yu_astr}.

So, let us perform averaging of the Einstein equations and the corresponding equations of physical fields. Let
us pay attention to the following circumstance. To begin the averaging, we need to have a certain <<primeval>> starting macroscopic metric $g^{(0)}_{ik}(x)$ and certain primeval starting physical fields $\phi^{(0)}_a$, which from we can start the process of consecutive averaging
iterations as a zero-order approximation. These primeval fields satisfy a self-consistent set of field equations:
\begin{eqnarray}\label{Einst0}
G^i_k\bigl(g_{jm}^{(0)}\bigr)=8\pi T^i_k(\bigl(g_{jm}^{(0)},\phi^{(0)}_a\bigr)+\Lambda \delta^i_k;\\
\label{Field0}
\pert{\nabla}{0}_k T^k_i(g^{(0)},\phi^{(0)}_a)=0.
\end{eqnarray}

So, in accordance with the common approach to averaging of gravitational fields, let us consider a certain macroscopic Riemannian space and let now  $g^{(0)}_{ik}(x)\equiv \overline{g_{ik}(x)}$, $\phi^{(0)}_a(x) \equiv \overline{\phi_{a}(x)}$ are certain {\it yet
unknown} macroscopic averages of the field values.

\begin{assume}\label{[d,int])=0}
Let us further assume that operations of differentiation or integration over coordinates commute with the averaging operation:
\begin{eqnarray}\label{[diff,average]=0}
\overline{\partial_i\phi(x)}=\partial_i\overline{\phi(x)};\\
\label{[int,average]=0}
\overline{\int \phi(x)dx}=\int\overline{\phi(x)}dx  .
\end{eqnarray}
\end{assume}
Let us notice that the above-defined operation of averaging completely satisfies the assumption \ref{[d,int]=0}. Let it be that:
\begin{equation}\label{dg,df}
\delta g_{ik}=g_{ik}-g^{(0)}_{ik};\quad \delta\phi_a=\phi_a-\phi^{(0)}_{a},
\end{equation}
are small local deviations of the metric and physical fields from the averaged values, so that:
\begin{eqnarray}
\label{aver_g}
\overline{\delta g_{ik}}=0; & \overline{\partial_j\delta g_{ik}}=0;\\
\label{aver_phi}
\overline{\delta \phi}=0; & \overline{\partial_j \delta\phi}=0.
\end{eqnarray}
Let us expand the field equations in Taylor series over the smallness of the metric and physical fields deviations from the average values up to the
second order in perturbations:
\begin{eqnarray}\label{Einst0-2}
G^{(0)}\ \!\!^i_k+G^{(1)}\ \!\!^i_k+G^{(2)}\ \!\!^i_k=8\pi\bigl( T^{(0)}\ \!\!^i_k+T^{(1)}\ \!^i_k+T^{(2)}\ \!\!^i_k\bigr)+\Lambda \delta^i_k,
\end{eqnarray}
and let us average now these equations with the account of (\ref{bar_dg=0}), (\ref{bar_df=0}), as well as linearity of the operators $G^{(1)}_{ik}$ and $T^{(1)}_{ik}$
with respect to fluctuations:
\begin{eqnarray}\label{G1=0}
\overline{G^{(1)}\ \!\!^i_k(\delta g)}=G^{(1)}\ \!\!^i_k(\overline{\delta g})=0;\\
\label{T1=0}
\overline{T^{(1)}\ \!\!^i_k(\delta g,\delta\phi)}=T^{(1)}\ \!\!^i_k(\overline{\delta g},\overline{\delta\phi})=0.
\end{eqnarray}
Thus, in the first order of perturbation theory, we get the microscopic linear equations of the first order in the smallness of metric's and physical fields'
perturbations
\begin{eqnarray}\label{evolut_dg}
G^{(1)}\ \!\!^i_k(\delta g)=8\pi T^{(1)}\ \!\!^i_k(\delta g,\delta\phi)+\Lambda \delta^i_k;\\
\label{evolut_df}
\nabla_k T^{(1)}\ \!\!^i_k(\delta g,\delta\phi)=0,
\end{eqnarray}
which we will call the {\it microscopic evolutionary equations for perturbations}. Assuming that\footnote{Let us notice that the substitution (\ref{gik=gik0}) is a formal technique for metric renormalization.}
\begin{equation}\label{gik=gik0}
g^{(0)}_{ik}=\overline{g_{ik}},
\end{equation}
in the second order of perturbation theory after the averaging we obtain the macroscopic Einstein equations for the macroscopic metric
\begin{eqnarray}\label{Eq_<Einst2>}
G^{(0)}\ \!\!^i_k(\overline{g})=-\overline{G^{(2)}\ \!\!^i_k(\delta g)}+8\pi T^{(0)}\ \!\!^i_k(\overline{g},\overline{\phi})
+8\pi\overline{T^{(2)}\ \!\!^i_k(\delta g,\delta\phi)} +\Lambda \delta^i_k,
\end{eqnarray}
according to which the macroscopic metric in the second order as per the theory of perturbations is determined by the Einstein equations with a cosmological
constant and the summed effective energy-momentum tensor:
\begin{equation}\label{tilde_MET}
\tilde{T}^i_k=T^{(0)}\ \!\!^i_k(\overline{g},\overline{\phi})+\mathcal{T}^{(2)}\ \!\!^i_k,
\end{equation}
where
\begin{equation}\label{Tik_sum}
\mathcal{T}^{(2)}\ \!\!^i_k\equiv\overline{ T^{(2)}\ \!\!^i_k}-\frac{1}{8\pi}\overline{G^{(2)}\ \!\!^i_k}.
\end{equation}
Thus, the macroscopic Einstein equations of the second order over perturbations take the following standard form:
\begin{equation}\label{Eq_Einst_aver}
G^i_k(\overline{g})=8\pi \tilde{T}^i_k +\Lambda \delta^i_k.
\end{equation}
For a closure of the macroscopic Einstein equations, it is required to calculate the macroscopic averages which are quadratic over the local fluctuations of
the metric and physical fields. These averages are determined by the evolution equations (\ref{evolut_dg}) -- (\ref{evolut_df}).

\subsection{The Macroscopic Symmetries}

The following object is called a Lie derivative of $\Omega^i_k$ in the direction of the vector field $\xi^i$ \cite{Petrov}.
\begin{equation}\label{Lee-det}
\Lee{\xi}U^i_k=\lim\limits_{dt\to 0}\frac{\Omega^i_k(x^i+\xi^i dt)-\Omega^i_k(x^i)}{dt},
\end{equation}
so that:
\begin{equation}\label{Lee_det}
\Lee{\xi}U^i_k=\xi^j\partial_j U^i_k-U^j_k\partial_j \xi^i+U^i_j\partial_k \xi^j.
\end{equation}
Particularly, if $U^i_k$ is a tensor and it is possible to replace partial derivatives in (\ref{Lee_det}) with covariant ones:
\begin{equation}\label{Lee_det_cov}
\Lee{\xi}U^i_k=\xi^j\nabla_j U^i_k-U^j_k\nabla_j \xi^i+U^i_j\nabla_k \xi^j
\end{equation}
-- this object is a tensor of the same valency as the source one.

Let us consider a pseudo-Riemannian space $V_4$, as the primeval space which allows a certain group of motions, $G^r$ with Killing vectors $\stackunder{(\alpha)}{\xi}$, determining a macroscopic symmetry::
\begin{eqnarray}\label{Lg0}
\Lee{\alpha}g^{(0)}_{ik}=0;\quad  \Lee{\alpha}\phi^{(0)}_{a}=0; \qquad   \bigl(\Lee{\alpha}\equiv \Lee{\xi_\alpha}\bigr).
\end{eqnarray}

Let us make the following assumption
\begin{assume}\label{as2}
The macroscopic average of the Einstein tensor inherits the symmetry properties of the macroscopic metric:
\begin{equation}\label{macr_sym}
\Lee{\xi}\overline{g_{ik}}=\sigma \overline{g_{ik}} \Longrightarrow \Lee{\xi} \overline{G_{ik}}= \sigma_1 \overline{g_{ik}},
\end{equation}
where $\Lee{\xi}$ is a Lie derivative (see, e.g., \cite{Petrov}), and $\sigma(x),\sigma_1(x)$ are certain scalar functions.
\end{assume}

Let us notice that at $\sigma=0,\ \sigma_1=0$ we get a group of motions of $\bar{V}_4$, while at nonzero values of these scalars we get a group of conformal transformations. Let us explain the assumption \ref{as2}. As is known from Riemannian geometry (see \cite{Yubook1}), all
geometric objects inherit the symmetry properties of the metric tensor, for instance:
\begin{equation}\label{Lg}
\Lee{\xi}g_{ik}=0 \Rightarrow \Lee{\xi}\Gamma^j_{ik}=0;\; \Lee{\xi}R_{ijkl}=0;\; \Lee{\xi}T_{ik}=0.
\end{equation}
Hence it follows, for example, that all symmetric covariant tensors of the second valency have the same algebraic structure as the metric tensor. Thus
it is logical to assume that the algebraic structure of macroscopic tensors will also be the same. Let us notice that since the following equalities are fulfilled:
\begin{equation}\label{LeeT0}
\Lee{\alpha}G^{(0)}\ \!^i_k=0; \quad \Lee{\alpha}\Lambda \delta^i_k=0; \quad\Lee{\alpha}T^{(0)}\ \!^i_k(\overline{g},\overline{\phi})=0,
\end{equation}
then, as a consequence of Assumption \ref{as2}, the summary energy-momentum tensor of the second order over perturbations should also inherit the symmetries
of the macroscopic metrics at averaging:
\begin{equation}\label{LeeT2=0}
\Lee{\alpha}\overline{\mathcal{T}^{(2)}\ \!^i_k}=0\Rightarrow \Lee{\alpha}\tilde{T}^i_k=0.
\end{equation}
Let us notice that Assumption \ref{as2} imposes certain necessary conditions on the symmetry of a scalar distribution function of random tensor
fields $f(x^i,\vec{\lambda})$, in particular:
\begin{equation}\label{f_sym}
\Lee{\alpha}f(x^i,\vec{\lambda})=0.
\end{equation}
\subsection{The Macroscopic Einstein Equations of the Second Order}

Let us finally write out a set of equations defining the macroscopic metrics in the second order of the perturbation theory. These equations consist
of linear equations for local perturbations of the metrics and physical fields
\begin{eqnarray}\label{Einst1}
G^{(1)}\ \!\!^i_k(\delta g)=8\pi T^{(1)}\ \!\!^i_k(\delta g,\delta\phi);\\
\label{Field1}
\nabla_k T^{(1)}\ \!\!^i_k(\delta g,\delta\phi)=0
\end{eqnarray}
and macroscopic Einstein equations
\begin{eqnarray}\label{Eq2<Einst2>}
G\ \!\!^i_k(\overline{g})-\Lambda \delta^i_k=-\overline{G^{(2)}\ \!\!^i_k(\delta g)}+
8\pi(T^{(0)}\ \!\!^i_k(\overline{g},\overline{\phi})+\overline{T^{(2)}\ \!\!^i_k(\delta g,\delta\phi)}),
\end{eqnarray}
where the values $G^i_k(\overline{g})$ are calculated with respect to the macroscopic metrics $\overline{g}_{ik}$.

Let us make the following important remark. Obtaining the macroscopic Einstein equations, we have performed a renormalization of the metric $g^{(0)}_{ik}\to
\overline{g}_{ik}$. There are some consequences of this fact. First of all, according to the self-consistent field method, the macroscopic averages $\overline{G^{(2)}_{ik}(\delta g)}$ are not necessarily small as compared to $G_{ik}(\overline{g})$, since they are not a result of summing of an infinite number of degrees of freedom of gravitational perturbations. Second, the cited renormalization of the metric leads to a change $G_{ik}(g^{(0)})\to G_{ik}(\overline{g})$. This, in turn, means that the macroscopic Einstein equations (\ref{Eq2<Einst2>}) are immediately resolved with respect
to the macroscopic metric and not by the method of consecutive iterations:
$$\overline{g_{ik}}=g^{(0)}_{ik}+\overline{\delta g_{ik}}+\ldots$$
-- this is one of the main advantages of the self-consistent field method of Hartree-Fock-Vlasov-Bogolyubov. The method of consecutive iterations
would have pulled us in a completely different direction. To understand that, it is enough to imagine the Einstein equations for the Friedmann Universe
where the energy-momentum tensor would have been defined by a gas of an infinite number of <<small photons>>. Following the method of consecutive
approximations, we would take the Minkowski tensor as the zero-order approximation. It is clear that we would never get a cosmological singularity
with that approach, taking into account the smallness of gravitational perturbations brought by <<small photons>>, as compared to units of the
Minkovsky metrics.  This example pictorially shows that all the researchers involuntarily use the self-consistent field method in the relativistic theory of
gravitation, not making attempts to justify it.

\section{The Core Relations of the Model}
\subsection{Field Equations}
To implement the above described averaging method let us consider the self-consistent system of Einstein equation of the classical scalar field $\Phi$ with Higghs potential as a certain field model. The following Lagrangian function corresponds to scalar field $\Phi$:
\begin{equation}\label{Ls}
\mathrm{L}_s=\frac{1}{8\pi}\biggl(\frac{1}{2}g^{ik}\Phi_{,i}\Phi_{,k}-V(\Phi)\biggr),
\end{equation}
where $V(\Phi)$ -- is a potential energy of scalar field which takes the following form for the Higgs potential:
\begin{equation}\label{Higgs}
V(\Phi )=-\frac{\alpha }{4} \left(\Phi ^{2} -\frac{m^{2} }{\alpha } \right)^{2} ,
\end{equation}
$\alpha$ is a constant of self-action, $m$ -is a mass of scalar bosons.

Let us obtain the equation of the scalar field from the Lagrangian function \eqref{Ls} with a help of standard procedure:
\begin{equation}\label{Eq_s}
\Box\Phi+V'_\Phi=0.
\end{equation}
Then %
\begin{equation}\label{T_{iks}}
T^i_k=\frac{1}{8\pi}\bigl(\Phi^{,i}\Phi_{,k}-\frac{1}{2}\delta^i_k\Phi_{,j}\Phi^{,j}+\delta^i_k m^2V(\Phi)\bigr)
\end{equation}
-- is a tensor of the energy - momentum of the scalar field. Let us omit the permanent member in the Higgs potential \eqref{Higgs} since it leads to simple redefinition of the cosmological constant $\Lambda$.

The corresponding Einstein equations of the examined system have the following form\footnote{In this paper it is everywhere $G=\hbar=c=1$, the signature of metrics is $(-,-,-,+)$, Ricci tensor is found by convolution of the first and third indices.}:
\begin{eqnarray}\label{Eq_Einst}
G^i_k\equiv R^i_k-\frac{1}{2}R\delta^i_k-\Lambda\delta^i_k =8\pi T^i_k.
\end{eqnarray}
For the sake of simplicity, let us further call equations \eqref{Eq_s} and (\ref{Eq_Einst})  the {\it field equations}.

\subsection{The Gravitational Perturbations of the Isotropic Universe}
Let us write out the metrics with gravitational perturbations in the following form (see., e.g., \cite{Land_Field}):
\begin{eqnarray}
\label{Freed}
ds^2_0=a^2(\eta)(d\eta^2-dx^2-dy^2-dz^2);\\
\label{metric_pert}
ds^2=ds^2_0-a^2(\eta)h_{\alpha\beta}dx^\alpha dx^\beta;
\end{eqnarray}
paying attention to the conformal factor $-a^2(\eta)$ of covariant amplitudes of perturbations which disappears for all mixed components of perturbations $h^\alpha_\beta$.  The covariant perturbations of metrics are equal to:
\begin{equation}\label{dg}
\delta g_{\alpha\beta}=-a^2(\eta)h_{\alpha\beta}.
\end{equation}
Then:
\begin{eqnarray}\label{defh1}
 h^\alpha_\beta=h_{\gamma\beta}g^{\alpha\gamma}_0\equiv-\frac{1}{a^2}h_{\alpha\beta};\\
 \label{defh2}
h\equiv h^\alpha_\alpha\equiv  g^{\alpha\beta}_0h_{\alpha\beta}=
 -\frac{1}{a^2}(h_{11}+h_{22}+h_{33}).
\end{eqnarray}
Let us consider Laplace operator $\triangle$ on 3-dimensional conformally euclidian space $\mathrm{E}_3$ with metrics
\begin{equation}\label{dl2}
d\ell^2=a^2(\eta)(dx^2+dy^2+dz^2)
\end{equation}
and find eigenfunctions of the self-conjugated operator
\begin{equation}\label{laplas_func}
\triangle\psi\equiv \frac{1}{a^2}\triangle_0\psi= k^2\psi,
\end{equation}
where $\triangle_0$ is a Laplace operator defined on 3-dimensional euclidian space and a $k^2(\eta)$ is a certain \emph{scalar} function.
Thus, the following expressions are the eigenfunctions of the Laplace operator $\triangle$ \eqref{laplas_func}:
\begin{equation}\label{self_func}
\psi_\mathbf{n}(\mathbf{r},\eta)=\mathrm{e}^{i\mathbf{nr}},
\end{equation}
where $n_\alpha=\mathrm{const}$ and $\mathbf{k}=\mathbf{n}/a(\eta)$ is a wave vector so that
\begin{equation}\label{n}
k^2=-g_0^{\alpha\beta}n_\alpha n_\beta\equiv \frac{\mathbf{n}^2}{a^2(\eta)}.
\end{equation}
Further on, since each mode of perturbations should be real value, let us rewrite it in the following form:
\begin{equation}\label{h_mode}
h_{\alpha\beta}(\mathbf{r},\eta)=h_{\alpha\beta}(\mathbf{n},\eta)\mathrm{e}^{i\mathbf{nr}+i\psi_0}+
h^*_{\alpha\beta}(\mathbf{n},\eta)\mathrm{e}^{-i\mathbf{nr}-i\psi_0},
\end{equation}
where $^*$ stands for operation of complex conjugation and $\psi_0$ is an arbitrary constant  {\it phase} for each harmonic $\mathbf{n}$.
The introduced arbitrary phase of the perturbation mode reflects the fact that as a result of homogeneity of 3-dimensional space each mode of metric's perturbation should not depend on the choice of the Cartesian system of coordinates origin. Thus as a consequence of translation symmetry we would have written a separate harmonics in the form $\mathrm{e}^{-i\mathbf{n(r-r_0)}}$, what was actually done when we have put $\psi_0=-\mathbf{r}_0\mathbf{n}$.
Thus, in each time instant $\eta_0$ a separate independent mode of metric's perturbations is fully described by 3-dimensional tensor amplitudes $h_{\alpha\beta}(\mathbf{n},\eta_0)$, classification of which along the direction of space-like vector $\mathbf{n}$
and 3-dimensional Kronecker tensor  $\delta_{\alpha\beta}$ is provided in the work by Lifshitz \cite{Lifshitz}.

For transverse perturbations it is
\begin{equation}
\label{h(eta)}
h_{\alpha\beta}=e_{\alpha\beta}\bigl(S\mathrm{e}^{i\mathbf{nr}+i\psi_0}+\mathbb{CC}\bigr),
\end{equation}
where $\mathbb{CC}$ stands for complex conjugate magnitude; $S(\mathbf{n},\eta),S^*(\mathbf{n},\eta)$ -- is an amplitude of perturbations so that
\begin{eqnarray}
\label{perp}
h^\alpha_\beta n_\alpha=0;\\
\label{perp1}
h=0.
\end{eqnarray}
In the arbitrary Cartesian system of coordinates of 3-dimensional Euclidian space $E_3$ the polarization tensor $e_{\alpha\beta}$ in formula (\ref{defh2}) takes the form:
\begin{eqnarray}\label{e_ab}
e_{\alpha\beta}=2s_\alpha s_\beta+\frac{n_\alpha n_\beta}{n^2}-\delta_{\alpha\beta},\\
\label{orto3}
\mathbf{s}^2=1; \quad \mathbf{sn}=0,\quad \mathbf{n}^2=n^2.
\end{eqnarray}
One can easily validate that gauge condition (\ref{perp}) is automatically fulfilled. Thus, each mode of transverse gravitational perturbations is described by {\it numbers}: $\mathbf{n}=(n_1,n_2,n_3)$, $\mathbf{s}=(s_1,s_2,s_3)$, $\psi_0$. Conditions (\ref{orto3}) impose 2 couplings on these numbers.
Thus, each mode of transverse gravitational perturbations possesses 5 degrees of freedom over which we should perform averaging.

For vector perturbations of the metrics it is:
\begin{eqnarray}\label{vect}
h_{\alpha\beta}=V_\alpha n_\beta+V_\beta n_\alpha; \quad (\mathbf{nV})=0;\\
\label{ortov}
\mathbf{V}=\mathbf{v} \mathrm{e}^{i\mathbf{nr}+i\psi_0}+\mathbb{CC},
\end{eqnarray}
where $\mathbf{v}(\mathbf{n},\eta),\mathbf{v}(\mathbf{n},\eta)$ -- is an amplitude of vector perturbations. Each mode of vector perturbations is described by 5 degrees of freedom.

Similarly, for the longitudinal perturbation of the metrics it is:
\begin{eqnarray}\label{long}
h_{\alpha\beta}=\mathrm{e}^{i\mathbf{nr}+i\psi_0} \bigl(\lambda P_{\alpha\beta}+\mu Q_{\alpha\beta}\bigr) +\mathbb{CC},
\end{eqnarray}
where $\lambda(\mathbf{n},\eta),\lambda^*(\mathbf{n},\eta),\mu(\mathbf{n},\eta),\mu^*(\mathbf{n},\eta)$ -- are amplitudes of scalar perturbations
\begin{equation}
P_{\alpha\beta}=\frac{1}{3}\delta_{\alpha\beta}-\frac{n_\alpha n_\beta}{n^2};\quad Q_{\alpha\beta}=\frac{1}{3}\delta_{\alpha\beta}.
\end{equation}

Let us notice papers \cite{calibr1} -- \cite{calibr5} and other (see review \cite{calibr6}, 2018), dedicated to the development of gauge invariant theory of scalar, vector and tensor perturbations of 2nd order and higher in the world of Friedmann. The main idea of this theory is specification of such canonical field degrees of freedom of perturbations where it would be possible to split aforementioned modes of perturbations in higher degrees of freedom of perturbation theory and specification of physically observable values like energy - momentum etc., corresponding to the modes. The introduction of such papers was caused by the requirements of quantum theory of the early Universe. Let us emphasize, that \emph{the theory of macroscopic Einstein equations of the 2nd order over perturbations} being developed in the paper and from one hand, being classical one, does not require selecting out such pure field modes for calculation of the physically observable values corresponding to them, and from the other hand, being macroscopic theory of the second order (macroscopic averages of perturbations are equal to zero), it require the fulfilment of the gauge conditions only in the first order of perturbation theory for which the above cited expansion of the Lifshitz perturbations is valid.

\subsection{The Averaging of the Einstein Equations over the States of Mode of Metric's Perturbations}
As was noted earlier, each independent mode of perturbations on the background of Friedmann's space is described by 5 independent numbers. At given wave vector $\mathbf{n}$ the polarization vector of the gravitational wave has only one degree of freedom remaining. Practically, this is an angle of rotation of a unit vector in a plane which is orthogonal to the wave vector. The wave vector itself $\mathbf{n}$ carries 3 degrees of freedom. One another degree of freedom is connected to the oscillation phase which can be defined on the interval $[0,2\pi]$. Longitudinal perturbations are fully set by 4 numbers, $n$ and $\psi_0$, i.e. have 4 degrees of freedom. Accordingly, we will carry out an operation of the metric's averaging over 4 degrees of freedom, leaving the polarization vector of the transverse gravitational perturbations fixed.

We will not carry out procedure of averaging over the length of wave vector since it would require setting a spectrum of perturbations\footnote{This operation at given spectrum of perturbations can be carried out on the final stage.}.
As a result, the operation of averaging over arbitrary perturbations of a certain value $f(n,\mathbf{r})$ can be written in the form:
\begin{equation}\label{average}
\overline{f(n,\mathbf{r})}=\frac{1}{8\pi^2}\int\limits_0^{2\pi}d\psi_0\int\limits_{\Omega_n} f(\mathbf{n},\mathbf{r})d\Omega_n.
\end{equation}
where operation of averaging over directions of wave vector is reduced to calculation of the integral over 2-dimensional sphere of radius $n$. As a result of isotropy and homogeneity of the phase space we have the following relations:
\begin{eqnarray}\label{<n>}
\overline{\mathbf{n}}=0;& \displaystyle \overline{n^\alpha n^\beta}=\frac{1}{3}\mathbf{n}^2;
\end{eqnarray}

When calculating the averaging one should take into account an arbitrary character of metric's $h_{\alpha\beta}$ and scalar field's $\varphi$  perturbations as a result of which we have:
\begin{eqnarray}\label{<h>=0}
\overline{h_{\alpha\beta}} =0;\\
\label{<f>=0}
\overline{\varphi}=0.
\end{eqnarray}
As a consequence of isotropy of the macroscopic metrics and (\ref{<h>=0}), (\ref{<f>=0}) the averaged in such a way summary tensor of energy - momentum should be isotropic i.e., having a structure of ideal fluid's energy - momentum tensor \footnote{Denotations see in \cite{G&C16}}:
\begin{equation}\label{ideal}
\overline{\mathcal{T}^{(2)}_{ik}}=a^2(\overline{\mathcal{E}}+\overline{\mathcal{P}})\delta^4_i\delta^4_k-a^2\eta_{ik}\overline{\mathcal{P}},
\end{equation}
where $\eta_{ik}=\mathrm{Diag}(-1,-1,-1,+1)$,
\begin{eqnarray}\label{E_mid}
\overline{\mathcal{E}} =\frac{1}{a^2}\overline{\mathcal{T}^{(2)}_{44}}; \\
\label{P_mid}
\overline{\mathcal{P}} =\frac{1}{3a^2} \overline{\sum\limits_{\alpha=1}^3\mathcal{T}^{(2)}_{\alpha\alpha}}
\end{eqnarray}
are macroscopic energy density and pressure of the field gravitationally scalar matter.

Let us notice that since summary energy - momentum tensor of the gravitational and scalar field and their first derivatives are included in a second order, the average from the product should be understood as follows:
\begin{equation}\label{average_ab}
\overline{ab} =\overline{ab^*+a^*b}.
\end{equation}
Due to isotropy and homogeneity of the macroscopic metrics after such operation of averaging we can find macroscopic Einstein equations (\ref{Eq_<Einst2>}) of the second order over perturbations for it:
\begin{eqnarray}\label{Freed_macrE}
3\frac{a'^2}{a^2}=\mathcal{E}_0+\overline{\mathcal{E}};\\
\label{Freed_macrP}
\frac{a'^2}{a^2}-2\frac{a''}{a}=\mathcal{P}_0+\overline{\mathcal{P}}.
\end{eqnarray}
Let us notice that according to the method of self-consistent field in these equations the scale factor is a macroscopic value.

The potential of the scalar field should be presented in the following form:
\begin{equation}\label{df}
\Phi=\Phi_0(\eta)+\phi(\eta)\mathrm{e}^{i\mathbf{nr}+i\psi_0}+\phi^*(\eta)\mathrm{e}^{-i\mathbf{nr}-i\psi_0}.
\end{equation}
Let us further expand the Einstein tensor and tensor of energy - momentum in a series by smallness of amplitude of perturbations of
gravitational and matter fields  $S(\eta), V(\eta),\lambda(\eta), \mu(\eta), \phi(\eta), \delta\varepsilon, \delta u^\alpha$.
Using the isotropy of non-perturbed metrics as well as conditions of orthogonality  \eqref{orto3}and \eqref{vect},
it is convenient to introduce a local system of coordinates where it is:
\begin{equation}\label{loc_coord}
\mathbf{n}=n(0,0,1); \quad \mathbf{s}=(1,0,0);\quad \mathbf{v}=(0,v(\eta),0),
\end{equation}
where $\mathbf{s}$ - is a unit vector of transverse perturbation polarization. In this system of coordinates it is
\begin{eqnarray}\label{nz1}
 h_{11}=\biggl[-S(\eta)+\frac{1}{3}\lambda(\eta)+\frac{1}{3}\mu(\eta)\biggr]\mathrm{e}^{inz}+\mathbb{CC};\nonumber\\
h_{22}=\biggl[S(\eta)+\frac{1}{3}\lambda(\eta)+\frac{1}{3}\mu(\eta)\biggr]\mathrm{e}^{inz}+\mathbb{CC};\nonumber \\
\label{nz3}
h_{12}=h_{13}=0; \quad h_{23}=\upsilon(\eta)\mathrm{e}^{inz}+\upsilon^*(\eta)\mathrm{e}^{-inz};\nonumber\\
\label{nz2}
h_{33}=\biggl[-\frac{2}{3}\lambda(\eta)+\frac{1}{3}\mu(\eta)\biggr]\mathrm{e}^{inz}+\mathbb{CC};\nonumber\\
h=\mu(\eta)\mathrm{e}^{inz}+\mu^*(\eta)\mathrm{e}^{-inz}.\nonumber
\end{eqnarray}
%

\section{Expanding of the Field Equations over Perturbation Orders}
\subsection{Null Approximation}
Expanding the Einstein tensor over metric's perturbations, in a null approximation we find the following known expressions:
\begin{eqnarray}\label{Einst0}
G^{(0)}\ \!^1_1=G^{(0)}\ \!^2_2=G^{(0)}\ \!^3_3=2\frac{a''}{a^3}-\frac{a'^2}{a^4};\quad
\order{G}{0}{^4_4}=3\frac{a'^2}{a^2}.
\end{eqnarray}
For the equation of the scalar field (\ref{Eq_s}) we get:
\begin{equation}\label{Eq_S0}
\Phi_0 ''+2\frac{a'}{a}\Phi_0 '+a^2(m^2 \Phi_0-\alpha\Phi^3_0)=0.
\end{equation}
The nonzero components of the energy - momentum tensor of the scalar field are equal to:
\begin{eqnarray}\label{TikS0}
\order{T}{0}{^1_1}=\order{T}{0}{^2_2}=\order{T}{0}{^3_3}\equiv-\mathcal{P}_0=
-\frac{\Phi_0'^2}{16\pi a^2}+\frac{m^2\Phi^2_0}{16\pi}-\frac{\alpha\Phi^4_0}{32\pi};\nonumber\\
\order{T}{0}{^4_4}\equiv\mathcal{E}_0=\frac{\Phi_0'^2}{16\pi a^2}+\frac{m^2\Phi^2_0}{16\pi}-\frac{\alpha\Phi^4_0}{32\pi}.
\end{eqnarray}
Thus, in a null approximation by gravitational perturbations we find two independent Einstein equations
\begin{eqnarray}\label{EqEinst0_4}
3\frac{a'^2}{a^4}-\frac{\Phi_0'^2}{2a^2}-\frac{m^2\Phi^2_0}{2}+\frac{\alpha\Phi^4_0}{4} - \Lambda=0;\\
\label{EqEinst0_1}
2\frac{a''}{a^3}-\frac{a'^2}{a^4}+\frac{\Phi'^2_0}{2a^2}-\frac{m^2\Phi^2_0}{2}+\frac{\alpha\Phi^4_0}{4}-\Lambda=0.
\end{eqnarray}
The equations< \eqref{EqEinst0_4} and \eqref{EqEinst0_1} compose a system of equations of base cosmological model for the scalar field with Higgs potential.
Differentiating the equation \eqref{EqEinst0_4} over time variable and substituting the expression for $a''/a^3$ from the equation \eqref{EqEinst0_1} into the obtained equation, let us find at $\Phi_0\not\equiv 0$ the field equation.
As we have noted earlier, in accordance with the method of self-consistent field we will not solve the \emph{background equations} \eqref{Eq_S0}, \eqref{EqEinst0_4} and \eqref{EqEinst0_1}. Instead, we will append the averaged quadratic corrections to it, which are not small values.

\subsection{The Equations of the First Order -- Evolution Equations for Perturbations}

In the approximation linear over $S-V-\lambda-\mu-\phi$ we can get, first of all, the equation for perturbation of the scalar field\footnote{Equations for complex conjugate amplitudes do not differ from it.}
\begin{eqnarray}\label{Eq_S1}
\!\!\!\phi''+2\frac{a'}{a}\phi'+a^2\phi\biggl(m^2+\frac{n^2}{a^2}-3\alpha\Phi^2_0\biggr)+\frac{\mu'}{2}\Phi_0'=0,
\end{eqnarray}
second, we can get nonzero components of the Einstein tensor $\order{G}{1}{^1_1}, \order{G}{1}{^2_2}, \order{G}{1}{^3_3}$, $\order{G}{1}{^3_2}=\order{G}{1}{^2_3}$, $\order{G}{1}{^1_4}=-\order{G}{1}{^4_1}$,  $\order{G}{1}{^2_4}=-\order{G}{1}{^4_2}$,
$\order{G}{1}{^3_4}=- \order{G}{1}{^4_3}$,and $\order{G}{1}{^4_4}$ and third, corresponding to them components of tensor energy - momentum. Separating the harmonics in the obtained equations $\mathrm{e}^{\pm ikz}$, let us write out independent combinations of corresponding non-trivial equations:
\begin{eqnarray}
\label{24}
\mbox{\normalsize ${4\choose2}:$} & \displaystyle v'=0;\\
\label{34}
\mbox{\normalsize ${4\choose3}:$} & \displaystyle \frac{1}{3}(\lambda+\mu)'+\phi\Phi'_0=0;\\
\label{22-11}
\mbox{\normalsize ${2\choose2}\!\!-\!\!{1\choose1}:$} & \displaystyle S''+2 \frac{a'}{a}S'+n^2S=0;\\
\label{2*33-11-22}
\mbox{\normalsize $2{3\choose3}\!\!-\!\!{2\choose2}\!\!-{1\choose1}\!\!:$} & \displaystyle\lambda''+2\frac{a'}{a}\lambda' -\frac{n^2}{3}(\lambda+\mu)=0;\\
\label{11+22+33}
\mbox{\normalsize ${1\choose1}\!\!+\!\!{2\choose2}\!\!+{3\choose3}\!\!:$} & \displaystyle \mu'' +2\frac{a'}{a}\mu'+\frac{n^2}{3}(\lambda+\mu)
 +3\phi'\Phi'_0-3a^2\Phi_0\phi(m^2-\alpha\Phi^2_0)=0\\
\label{44}
\mbox{\normalsize ${4\choose4}:$} & \displaystyle \frac{a'}{a}\mu'+\frac{n^2}{3}(\lambda+\mu)- \Phi'_0\phi'-a^2\phi\Phi_0(m^2-\alpha\Phi^2_0)=0.
\end{eqnarray}
The equation {\normalsize ${2\choose3}$} is a differential consequence of the equation \eqref{24}. Let us write out the equation ${3\choose3}$, which is an algebraic consequence of equations \eqref{2*33-11-22} and \eqref{11+22+33} (a one-third fraction of its sum)
\begin{equation}\label{33}
\mbox{\normalsize ${3\choose3}:$}\quad (\lambda+\mu)''+2\frac{a'}{a}(\lambda+\mu)'+3\Phi'_0\phi'-3a^2\Phi_0\phi(m^2-\alpha\Phi^2_0)=0.
\end{equation}
So, the equations \eqref{24} and \eqref{22-11} define amplitudes of vector and tensor perturbations. The remaining four equations \eqref{34}, \eqref{2*33-11-22}, \eqref{11+22+33} and \eqref{44} define four functions: amplitude of scalar perturbations of metrics $\lambda, \mu$ and amplitude of perturbation of the scalar field $\phi$, herewith equation \eqref{2*33-11-22} does not depend on the scalar field and fully coincide with the equation of Lifshitz theory.
Let us notice that one can use the consequences of equations of null approximation in the evolution equations since the subsequent terms will have a third order of perturbations. Let us also notice that from all the conservation laws $\nabla_kT^i_k=0$ of the first order there are only two which are nontrivial: $\nabla_kT^i_3=0$ and $\nabla_k T^k_4=0$, where the first one is equivalent to the equation of field of null approximation \eqref{Eq_S0}, while the second is equivalent to the field equation of the first order of theory of perturbations \eqref{Eq_S1}. Therefore with the account of equation for perturbations of the scalar field \eqref{Eq_S1}, from four equations \eqref{34}, \eqref{2*33-11-22}, \eqref{11+22+33} and \eqref{44} there are only two which are independent. Therefore one can select out in a capacity of base equations on three functions $\mu,\lambda,\phi$ for instance, a subsystem of equations  \eqref{34}, \eqref{2*33-11-22}f and \eqref{44}.

Further on we designate the system of linear equations \eqref{24},  \eqref{34}, \eqref{22-11}, \eqref{2*33-11-22} and \eqref{44} \emph{evolution equations} for perturbations of gravitational and scalar fields. The solutions of these equations define the quadratic corrections to the energy - momentum tensor and that way, the macroscopic metric of the Universe.
\subsection{The Approximation of the Second Order}
Let us calculate now the second order corrections to the Einstein tensor and energy - momentum tensor in accordance with \eqref{Tik_sum}:
\begin{equation}\label{G2-T2}
\mathcal{G}^{(2)i}_{~~~k}=G^{(2)i}_{~~~k}-8\pi T^{(2)i}_{~~~k}\equiv -8\pi \mathcal{T}^{(2)i}_{~~~k}.
\end{equation}
When calculating these corrections one should account equations of null and first orders over perturbations. Calculating, let us find:
\begin{equation}\label{GT2=0}
\mathcal{G}^{(2)1}_{~~~2}=\mathcal{G}^{(2)1}_{~~~3}=\mathcal{G}^{(2)1}_{~~~4}=\mathcal{G}^{(2)2}_{~~~4}=0.
\end{equation}
Next, the remaining nonzero component is $\mathcal{G}^{(2)4}_3\sim n_3\ (\mathbf{n})$, therefore as a consequence of \eqref{<n>} it is
\begin{equation}\label{GT^4_a}
\overline{\mathcal{G}^{(2)4}_{~~~\alpha}}=0.
\end{equation}
Besides, at calculation of mean-square corrections to energy - momentum tensor the terms containing $\mathrm{e}^{\pm 2i\mathbf{kr}}$ turn to null in consequence of isotropy of 3-dimensional space, therefore only summands which do not contain these factors are essential in formulas for corrections to energy and pressure \eqref{E_mid} and \eqref{P_mid}. With the account of these comments and evolution equations on amplitudes of perturbations \eqref{Eq_S1} -- \eqref{33}, performing averaging we obtain corrections to background pressure $\mathcal{P}_0$ and energy density $\mathcal{E}_0$ \eqref{TikS0}:
\begin{eqnarray}\label{P_g}
\overline{\mathcal{P}}=\frac{1}{24\pi}\overline{\sum\limits_{\alpha=1}^3 \mathcal{G}^{(2)\alpha}_{~~~\alpha}}=\frac{1}{8\pi}\biggl[
\frac{7n^2}{6a^2}\overline{SS^*}-\frac{5}{6a^2}\overline{S'S'\ \!^*} -\frac{n^2}{9}\biggl(\frac{7}{6}\overline{\lambda\lambda^*}+\frac{2}{3}\overline{(\lambda\mu^*+\lambda^*\mu)} -\frac{1}{6}\overline{\mu\mu^*} \biggr)\nonumber\\ -\frac{5}{18}\overline{\lambda'\lambda'\ \!^*}+\frac{1}{18}\overline{\mu'\mu'\ \!^*}
-\overline{\frac{\phi'\phi^*\ \!'}{a^2}}-\overline{\phi\phi^*}\biggl(\frac{n^2}{3a^2}+m^2-3\alpha\Phi^2_0\biggr)\biggr].\\
\label{E_g}
\overline{\mathcal{E}}=-\overline{\mathcal{G}^{(2)4}_{~~~4}}=\frac{1}{8\pi}\biggl[
n^2\biggl(\frac{\overline{SS^*}}{2a^2}+\frac{\overline{\lambda\mu^*+\lambda^*\mu}}{9}-\frac{\overline{\lambda\lambda^*}}{18}
\biggr)+\frac{\overline{S'S'\ \!^*}}{2a^2}+\frac{\overline{\lambda'\lambda'\ \!^*}}{6}-\frac{\overline{\mu'\mu'\ \!^*}}{6}\nonumber\\
+2\frac{a'}{a^3}\overline{(SS^*)'}+\frac{1}{3}\frac{a'}{a}\overline{(\mu\mu^*)'}+\frac{\overline{\phi'\phi^*\ \!'}}{a^2}
+\overline{\phi\phi^*}\biggl(\frac{n^2}{a^2}+m^2-3\alpha\Phi^2_0\biggr)\biggr].
\end{eqnarray}

\subsection{The Equations of Macroscopic Scalar Field of the Second Order over Perturbations and Macroscopic Einstein Equations}
To build a closed theory of macroscopic Universe it is necessary to obtain the equations of the second order over perturbations for the scalar field $\Phi$ on the basis of field equation \eqref{Eq_s}. Calculating the quadratic correction to the field equation we find:
\begin{eqnarray}\label{Eq_S0}
\delta^{(2)}(\Box\Phi+V'_\Phi)=-\Phi'_0\biggl[\frac{(SS^*)'}{a^2}+\frac{1}{3}(\lambda\lambda^*)'+\frac{1}{6}(\mu\mu^*)'\biggr]\nonumber\\
+\frac{1}{2}\frac{\phi'\mu'\ \!^*+\phi'\ \!^*\mu'}{a^2}-\frac{1}{2}\frac{n^2}{a^2}(\phi\mu^*+\phi^*\mu)-6\alpha\Phi_0\phi\phi^*.
\end{eqnarray}
Thus, according to the method of self-consistent field to define macroscopic scalar field $\Phi_0(\eta)$ instead of equation \eqref{Eq_S0} we have:
\begin{eqnarray}\label{d2Eq_s}
\Phi_0 ''+2\frac{a'}{a}\Phi_0 '+a^2(m^2\Phi_0-\alpha\Phi^3_0)- \Phi'_0 a^2\biggl[\frac{\overline{(SS^*)'}}{a^2}+\frac{1}{3}\overline{(\lambda\lambda^*)'}+\frac{1}{6}\overline{(\mu\mu^*)'}\biggr]\nonumber\\
+\frac{1}{2}\overline{\phi'\mu'\ \!^*+\phi'\ \!^*\mu'}- \frac{1}{2}n^2\overline{(\phi\mu^*+\phi^*\mu)}-6a^2\alpha\Phi_0\overline{\phi\phi^*}=0.
\end{eqnarray}
Proceeding similarly with Einstein equations \eqref{EqEinst0_4} with the account of \eqref{P_g} and \eqref{E_g} let us find instead of them the macroscopic Einstein equations:
\begin{eqnarray}\label{EqEinst_Macr_4}
3\frac{a'^2}{a^4}-\frac{\Phi_0'^2}{2a^2}-\frac{m^2\Phi^2_0}{2}+\frac{\alpha\Phi^4_0}{4} - \Lambda =n^2\biggl(\frac{\overline{SS^*}}{2a^2}+\frac{\overline{\lambda\mu^*+\lambda^*\mu}}{9}-\frac{\overline{\lambda\lambda^*}}{18}
\biggr)+\frac{\overline{S'S'\ \!^*}}{2a^2}+\frac{\overline{\lambda'\lambda'\ \!^*}}{6}-\frac{\overline{\mu'\mu'\ \!^*}}{6}\nonumber\\
+2\frac{a'}{a^3}\overline{(SS^*)'}+\frac{1}{3}\frac{a'}{a}\overline{(\mu\mu^*)'}+\frac{\overline{\phi'\phi^*\ \!'}}{a^2}
+\overline{\phi\phi^*}\biggl(\frac{n^2}{a^2}+m^2-3\alpha\Phi^2_0\biggr);\\
\label{EqEinst_Macr_1}
2\frac{a''}{a^3}-\frac{a'^2}{a^4}+\frac{\Phi'^2_0}{2a^2}-\frac{m^2\Phi^2_0}{2}+\frac{\alpha\Phi^4_0}{4}-\Lambda+
\frac{7n^2}{6a^2}\overline{SS^*}-\frac{5}{6a^2}\overline{S'S'\ \!^*} -\frac{n^2}{9}\biggl(\frac{7}{6}\overline{\lambda\lambda^*}+\frac{2}{3}\overline{(\lambda\mu^*+\lambda^*\mu)} -\frac{1}{6}\overline{\mu\mu^*} \biggr)\nonumber\\ -\frac{5}{18}\overline{\lambda'\lambda'\ \!^*}+\frac{1}{18}\overline{\mu'\mu'\ \!^*}
-\overline{\frac{\phi\phi'}{a^2}}-\overline{\phi\phi^*}\biggl(\frac{n^2}{3a^2}+m^2-3\alpha\Phi^2_0\biggr)=0.
\end{eqnarray}

The system of linear evolution equations \eqref{24},  \eqref{34}, \eqref{22-11}, \eqref{2*33-11-22} and \eqref{44} together with the system of macroscopic equations \eqref{d2Eq_s}, \eqref{EqEinst_Macr_4} and \eqref{EqEinst_Mecr_1} comprise a complete closed system of equations defining the macroscopic Friedmann Universe.

\subsection{Some Considerations on Macroscopic Equations}
\qquad Let us notice first that vector perturbations do not influence on macroscopic equation of scalar field and macroscopic Einstein equatinos, therefore further on we assume local vector perturbations are equal to zero, accounting that their linear equations \eqref{24} are autonomous.

Second, summing up equations \eqref{2*33-11-22} and \eqref{11+22+33} and substituting the expression for $(\lambda+\mu)'$  from \eqref{34} into the result, we get:
\begin{equation}\label{EqPhi_0}
\phi\biggl[\Phi''+2\frac{a'}{a}\Phi'+a^2\Phi\bigl(m^2-\alpha\Phi^2\bigl)\biggl]=0.
\end{equation}
The expression in square brackets \eqref{EqPhi_0} coincides with the left part of non-perturbed equation for background scalar field
\begin{equation}\label{Eq_Phi0_0}
\Phi_0 ''+2\frac{a'}{a}\Phi_0 '+a^2(m^2\Phi_0-\alpha\Phi^3_0)=0.
\end{equation}
The difference of this equation from the macroscopic equation for background scalar field, has second order of smallness over perturbations (see \eqref{d2Eq_s}). Therefore left side of the equation \eqref{EqPhi_0} has a third order over perturbations, consequently, one can assume the expression in square brackets \eqref{EqPhi_0} to be equal to zero in the theory of perturbations of the second order. Since the equation \eqref{EqPhi_0} was obtained as differential -- algebraic consequence of equations  \eqref{34}, \eqref{2*33-11-22} and \eqref{11+22+33}, one of these equations in the framework of quadratic over perturbations approximation can be disposed of. Thus, we consider the following system of four evolution equations: \eqref{Eq_phi}, \eqref{22-11}, \eqref{2*33-11-22} and \eqref{11+22+33} with respect to four functions: $\phi(\eta)$, $S(\eta)$, $\lambda(\eta)$ and $\mu(\eta)$.

Finally, let us notice also the peculiarity of equations of the macroscopic model mentioned in the introduction, namely the fact that linear evolution equations with respect to perturbations $\phi(\eta)$, $S(\eta)$, $\lambda(\eta)$ and $\mu(\eta)$ of scalar field and Friedmann metric \eqref{Eq_phi}, \eqref{22-11}, \eqref{2*33-11-22} and \eqref{11+22+33} are defined on the background of macroscopic values $\Phi(t),a(t)$, whhich are in turn, defined through nonlinear equations \eqref{d2Eq_s} -- \eqref{EqEinst_Macr_1} with the help of quadratic fluctuations of perturbations. This peculiarity of the equations of macroscopic model does not allow to obtain their solution by direct methods. Corresponding evolution equations in the standard theory of perturbations are being solved on the \emph{given} metric  $a(\eta)$. In our case we have to solve these equations at not determined background functions. Therefore it is necessary to solve both evolution equations and macroscopic ones simultaneously making assumptions regarding background functions $a(\eta)$ and $\Phi(\eta)$ nd perturbations $\phi(\eta)$, $S(\eta)$, $\lambda(\eta)$ and $\mu(\eta)$. One of such assumptions is the assumption about characteristic scales of change of these functions which are defined by wave number $n$ and characteristic dimension of background's inhomogeneity $f$: $\ell^{-1}=f'/f$.

\subsection{Consistency of the System of Equations of the Macroscopic Model}
\qquad The system of macroscopic equations consists of three differential equations of the 2nd order \eqref{d2Eq_s} -- \eqref{EqEinst_Macr_1} with respect to two macroscopic functions $a(\eta)$ and $\Phi(\eta)$. Therefore first of all one should investigate this system on its consistency. From the Einstein equations' standpoint the problem of consistency of this system of equations is connected with fulfilment of Bianchi identity laws for the Einstein tensor. In the case of isotropic and homogenous metric differential Bianchi identities are reduced to one relation. Let us prove consistency of the obtained equations.

In order to avoid usage of cumbersome formulas let us write down the system \eqref{d2Eq_s} -- \eqref{EqEinst_Macr_1} in the form:
\begin{eqnarray}\label{Macr_field}
\Phi''+2\frac{a'}{a}\Phi' +a^2\Phi(m^2-\alpha\Phi^2)=F;\\
\label{MacrEinst_44}
3\frac{a'^2}{a^4}-\frac{\Phi'^2}{2a^2}-\frac{m^2\Phi^2}{2}+\frac{\alpha\Phi^4}{4} - \Lambda =G;\\
\label{MacrEinst_11}
2\frac{a''}{a^3}-\frac{a'^2}{a^4}+\frac{\Phi'^2}{2a^2}-\frac{m^2\Phi^2}{2}+\frac{\alpha\Phi^4}{4}-\Lambda=H,
\end{eqnarray}
where functions $F,G,H$ are defined by mean-square perturbations of metrics and scalar field:
\begin{eqnarray}
F=& \displaystyle \Phi'\biggl[\overline{(SS^*)'}+\frac{1}{3}\overline{(\lambda\lambda^*)'}
+\frac{1}{6}\overline{(\mu\mu^*)'}\biggr]+6a^2\alpha\overline{\phi\phi^*}
 -\frac{1}{2}\overline{\phi'\mu'\ \!^*+\phi'\ \!^*\mu'}+ \frac{1}{2}n^2\overline{(\phi\mu^*+\phi^*\mu)};\nonumber\\
G=&  \displaystyle  n^2\biggl(\frac{\overline{SS^*}}{2a^2}+\frac{\overline{\lambda\mu^*+\lambda^*\mu}}{9}-\frac{\overline{\lambda\lambda^*}}{18}
\biggr)+\frac{\overline{S'S'\ \!^*}}{2a^2}   +\frac{\overline{\lambda'\lambda'\ \!^*}}{6}-\frac{\overline{\mu'\mu'\ \!^*}}{6}
+2\frac{a'}{a^3}\overline{(SS^*)'}+\frac{1}{3}\frac{a'}{a}\overline{(\mu\mu^*)'}\nonumber\\
 &  \displaystyle  + \frac{\overline{\phi' \phi'\ ^*}}{a^2}+\overline{\phi\phi^*}\biggl(\frac{n^2}{a^2}+m^2-3\alpha\Phi^2\biggr);\nonumber\\
H=&  \displaystyle  n^2\frac{7}{6a^2}\overline{SS^*}-\frac{5}{6a^2}\overline{S'S'\ \!^*} -\frac{n^2}{9}\biggl(\frac{7}{6}\overline{\lambda\lambda^*}-\frac{1}{6}\overline{\mu\mu^*} +\frac{2}{3}\overline{(\lambda\mu^*+\lambda^*\mu)} \biggr) -\frac{5}{18}\overline{\lambda'\lambda'\ \!^*}+\frac{1}{18}\overline{\mu'\mu'\ \!^*}\nonumber\\
& \displaystyle  -\overline{\frac{\phi'\phi'\ ^*}{a^2}}-\overline{\phi\phi^*}\biggl(\frac{n^2}{3a^2}+m^2-3\alpha\Phi^2\biggr)\nonumber
\end{eqnarray}

Differentiating the equation \eqref{MacrEinst_44} and using in the obtained expression equations \eqref{Macr_field} and \eqref{MacrEinst_11},
we obtain the relation:
\begin{equation}\label{Dif_Cons}
3\frac{a'}{a}(H-G)-\frac{\Phi'}{a^2}F=G'.
\end{equation}
The relation \eqref{Dif_Cons} is necessary and sufficient \footnote{The sufficiency is elementary proved.} by consistency condition of macroscopic equations \eqref{d2Eq_s} -- \eqref{EqEinst_Macr_1}. In case of absence of local fluctuations  ($F=G=H=0$) this condition is identically fulfilled. With local perturbations present, to verify the correctness of the relation \eqref{Dif_Cons} it is required to perform direct calculation. Doing that, it is required, first of all, to take into account commutativity of differentiation, complex conjugation and averaging operations. Second, it is necessary to use the evolution relations \eqref{Dif_Cons}. We do not represent here the quite cumbersome calculations, however show on a simple example main techniques of calculation of such values.

Let us calculate, for instance, the value $(\overline{S'S'\ ^*})'$: $(\overline{S'S'\ ^*})'=\overline{S''S'\ \!^*}+ \overline{S'S''\ \!^*}.$
From the evolution equation \eqref{22-11} we find:
\[S''=-2\frac{a'}{a}S'-n^2S.\]
Thus, we get:
\begin{eqnarray}
(\overline{S'S'\ \!^*})'= -4\frac{a'}{a} \overline{S'S'\ \!^*}-n^2 \overline{SS'\ \!^*}-n^2\overline{S'S^*}
\Rightarrow (\overline{S'S'\ \!^*})'=-4\frac{a'}{a} \overline{S'S'\ \!^*}-n^2 (\overline{SS^*})'.\nonumber
\end{eqnarray}
Dealing the same way with the second derivatives of functions $\phi, \lambda$ and $\mu$, let us prove the correctness of condition \eqref{Dif_Cons}.

Thus, further on we will define the macroscopic cosmological model by evolution equations \eqref{Eq_phi}, \eqref{22-11}, \eqref{34}, \eqref{2*33-11-22} and \eqref{11+22+33} and macroscopic equations \eqref{d2Eq_s}, \eqref{EqEinst_Macr_4} .

\section{The Examples of Construction of the Macroscopic Models of the Universe}
\subsection{The Case of Purely Transverse Perturbations}
The obtained system of equations of the macroscopic cosmology is extremely complicated for the analysis within a single paper. We intend to come back to its research in the nearest future. Meanwhile, as an example of the investigation of this system we consider the case of absence of scalar and vector perturbations of the gravitational field $\mu=\lambda=v=0$. In this case from the equation \eqref{34} we obtain $\phi=0$, and from this system of evolution equations there remains just the equation for transverse perturbations \eqref{22-11}.
Thus, the equation \eqref{d2Eq_s} is reduced to simpler equation:
\begin{eqnarray}\label{d2Eq_s1}
\Phi_0 ''+\Phi'_0\bigl[2\frac{a'}{a}- \overline{(SS^*)'}\bigr]+a^2\Phi_0\bigl(m^2-\alpha\Phi^2_0\bigr)=0.
\end{eqnarray}
Thus, the mean-square correction in the macroscopic equation of the scalar field in this case influences only on the value of the Hubble constant $H=a'/a$ by means of energy of gravitational waves.
From  \eqref{P_g} and \eqref{E_g} we get the expressions for corrections to effective pressure and energy density:
\begin{eqnarray}\label{P_gS}
\overline{\mathcal{P}}=\frac{1}{8\pi a^2}\biggl(
\frac{7}{6}n^2\overline{SS^*}-\frac{5}{6}\overline{S'S'\ \!^*} \biggr).\\
\label{E_gS}
\overline{\mathcal{E}}=\frac{1}{8\pi a^2}\biggl[
n^2\frac{\overline{SS^*}}{2}+\frac{\overline{S'S'\ \!^*}}{2}+2\frac{a'}{a}\overline{(SS^*)'}\biggr].
\end{eqnarray}
\subsection{WKB - Approximation for Transverse Perturbations}
Let us not consider the WKB-approximation of the evolution equation \eqref{22-11}
\[n\gg \frac{a'}{a},\quad S'\gg \frac{a'}{a}, \]
representing the solution in the form
\[S=\tilde{S}(\eta) e^{i\int u(\eta)d\eta,}\]
where $\tilde{S}(\eta)$ and $u(\eta)$ are functions slightly changing together with the scale factor, such that it is:
\begin{equation}\label{WKB}
a'\ll a \{n,u\};\quad \tilde{S}'\ll \tilde{S} \{n,u\}; \quad u'\ll \{n,u\}.
\end{equation}
Thus, let us find in the WKB- approximation:
\begin{equation}\label{S_WKB}
S= \frac{1}{a}S^0_+ \mathrm{e}^{in\eta}+\frac{1}{a}S^0_- \mathrm{e}^{-in\eta},
\end{equation}
where $S^0_\pm$ are constant amplitudes so that $S^0_+S^0_-=|S^0|^2$. Thus, let us find in the WKB - approximation:
\[SS^*\simeq\frac{|S^0|^2}{a^2}\Rightarrow \overline{SS^*}\simeq\frac{|S^0|^2}{a^2}; \quad S'S'\ \!^* \simeq n^2 \frac{|S^0|^2}{a^2}\Rightarrow \overline{S'S'\ \!^*}\simeq n^2\frac{|S^0|^2}{a^2}; \quad (SS^*)'\simeq 0. \]
Using these relations in formulas \eqref{P_gS} -- \eqref{E_gS}, we find in the WKB - approximation:
\begin{eqnarray}\label{PE_gS_WKB}
\overline{\mathcal{P}}\simeq\frac{1}{24\pi}n^2 \frac{|S^0|^2}{a^4}; & \displaystyle \overline{\mathcal{E}}\simeq\frac{1}{8\pi}n^2\frac{|S^0|^2}{a^4}\Rightarrow \overline{\mathcal{P}}\simeq\frac{1}{3}\overline{\mathcal{E}}
\end{eqnarray}
-- i.e., the account of high frequency transverse gravitational perturbations is equivalent to addition of component of ultrarelativistic fluid in macroscopic Einstein equations  \cite{G&C16}.

Next, substituting the WKB - solution \eqref{S_WKB} in macroscopic equation of scalar field, let us reduce it to the explicit form:
\begin{eqnarray}\label{d2Eq_s2}
\Phi_0 ''+2\frac{a'}{a}\Phi'_0\biggl(1+\frac{|S^0|^2}{a^2}\biggr)+a^2\Phi_0\bigl(m^2-\alpha\Phi^2_0\bigr)=0.
\end{eqnarray}
Similarly let us reduce the independent Einstein equation to the explicit form
\begin{eqnarray}\label{EqEinst_Macr_4_2}
3\frac{a'^2}{a^4}-\frac{\Phi_0'^2}{2a^2}-\frac{m^2\Phi^2_0}{2}+\frac{\alpha\Phi^4_0}{4} - \Lambda =n^2\frac{|S^0|^2}{a^4}.
\end{eqnarray}
\subsection{The Specific Solution of the Einstein Macroscopic Equations}
To get the solutions of equations of the macroscopic gravitation it is required to, first of all, find the solution of macroscopic equation of the scalar field \eqref{d2Eq_s2}, which itself is challenging at unknown scale factor. Papers \cite{Ign_Kokh,Ign_Kokh2} contain examples of such researches allowing to bypass this principal difficulty. Below we consider a simple example of specific exact solution of the field equation \eqref{d2Eq_s2}, allowing to finalize the task resolution.
Actually, assuming $\alpha>0$, let us put in \eqref{d2Eq_s2}
\begin{equation}\label{F=F0}
\Phi_0=\pm \frac{m}{\sqrt{\alpha}},
\end{equation}
-- in this case the equation \eqref{d2Eq_s2} turns into identity and the Einstein equation \eqref{EqEinst_Macr_4_2} is reduced to explicit form:
\begin{equation}\label{EqEinst_Macr_4_3}
3\frac{a'^2}{a^4}- \lambda =n^2\frac{|S^0|^2}{a^4},
\end{equation}
where it is
\begin{equation}\label{L->l}
\lambda=\Lambda+\frac{m^4}{4\alpha}.
\end{equation}
The equation \eqref{EqEinst_Macr_4_3} is integrated in elementary functions:
\begin{equation}\label{a_sol}
a(t)=\biggl(\frac{|S^0|^2 n^2}{\lambda}\biggr)^{1/4}\sqrt{\sinh \biggl(2\sqrt{\frac{\lambda}{3}}t\biggr)}.
\end{equation}
Similar solution has been found by the Author for the simpler macroscopic model of the Universe with a cosmological term where the scalar field is absent \cite{G&C16}. The difference of the specific solution of the cosmological model with a scalar field \eqref{a_sol} from this case practically reduces to renormalization of the cosmological constant \eqref{L->l}.
\section*{The Conclusion}
Thus, the closed system of macroscopic Einstein - Higgs equations describing the evolution of the macroscopically homogenous and isotropic Universe filled with fluctuating scalar field with Higgs potential was obtained. This system contains from the subsystem of ordinary linear differential equations describing evolution of perturbations of gravitational and scalar fields, and system of nonlinear macroscopic equations describing macroscopic dynamics of the cosmological model.
Let us notice the following property of such macroscopic models which differentiate them from standard cosmological models with homogenous scalar fields $\Phi(t)$: the self-consistent solution \eqref{a_sol} always contains the cosmological singularity. Actually, at $t\to 0$ the solution \eqref{a_sol}
behaves itself as a solution for ultrarelativistic Universe $a\simeq (4n^2|S^0|^2/3)^{1/4}\sqrt{t}\to 0$, and at $t\to\infty$ -- as inflationary one $a\simeq (n^2|S^0|^2/4\lambda)^{1/4}\exp(\sqrt{\lambda/3}t)$.   Let us notice that self-consistent description of the macroscopic model changes the real cosmological scenario radically, removing infinite past of the Universe from the standard scenario. The Universe, from one hand, is a macroscopic classical object. From the other hand, on the early stages of its evolution, its local properties are defined by quantum processes. However, these quantum processes, in turn, run on the classical (massive) gravitational background, defined by macroscopic laws\footnote{The prevalent classical, inert character of the gravitational background maintained by the Bierhoff theorem, was noted in the Author's cited works.}.


%

\end{document}